\documentclass[doublespacing]{elsart}

\usepackage{amssymb}
\usepackage{graphicx}
\usepackage{amsmath}

\journal{Physics Letters A}
\begin{document}

\begin{frontmatter}

\title{Dixon-Souriau equations from a 5-dimensional spinning particle in a Kaluza-Klein framework}

\author[icra]{F. Cianfrani},
\ead{francesco.cianfrani@icra.it}
\author[icra]{I. Milillo},
\ead{milillo@icra.it}
\author[icra,enea]{G. Montani}
\ead{montani@icra.it}

\address[icra]
{ICRA-International Center for Relativistic Astrophysics\\
Dipartimento di Fisica (G9),
Universit\`a  di Roma, ``La Sapienza",\\
Piazzale Aldo Moro 5, 00185 Rome, Italy}

\address[enea]
{ENEA C.R. Frascati (Dipartimento F.P.N.),\\ via Enrico Fermi 45, 00044 Frascati, Rome, Italy}

\begin{abstract}
The dimensional reduction of Papapetrou equations is performed in a 5-dimensional Kaluza-Klein background and Dixon-Souriau results for the motion of a charged spinning body are obtained. The splitting provides an electric dipole moment, and, for elementary particles, the induced parity and time-reversal violations are explained. 
\end{abstract}


\end{frontmatter}

\section{Introduction}
Among other unification theories, Kaluza-Klein models are those that better pursue the spirit of Relativistic Theories, i.e. the geometrization of all fundamental interactions.\\
The original Kaluza work dealt with the Electrodynamics only \cite{K1,K2}, but, not before all interactions but gravity were recognized as gauge theories, extensions of this approach to the non-Abelian sector followed \cite{ACF87,W81}(for a review on this topic see \cite{OW97} or \cite{Io}). This approach relays on the interpretation of extra-dimensional degrees of freedom as internal ones in a 4-dimensional space-time. In particular, the identification between translations on the extra-space and gauge transformations, constraints the number of space-time dimensions which should be added to reproduce gauge generators. In this scheme, the different behavior between gravity and other interactions arises as a consequence of the features of the space-time manifold. In fact, it is assumed that such a manifold is the direct sum of a 4-dimensional space-time with a compact homogeneous space, whose compactification to Planckian scales accounts for its present undetectability. Here, gauge bosons come out as additional off-diagonal components of the metric.\\
Hence, the issue of geometrization requires that the motion of bodies has to be inferred from geometric properties of the space-time. The equations describing the dynamics of an object in a generic space-time manifold are due to Mathisson \cite{2} and Papapetrou \cite{1,3}. In their works, the dynamics of a body is represented by a multipole expansion. It consists in describing the motion of the system by a finite number of moments only. This kind of approximation remains valid if the object under investigation has a small size with respect to the characteristic length scale of the gravitational field. In particular, in the pole-dipole case, they recognized that the obtained equations are appropriate to describe the motion of a spinning particle.\\ 
At the same order of approximation, even though by a rather different approach, Dixon \cite{5} and Souriau \cite{4} generalized this result to the case of a charged body coupled to an external electro-magnetic field.\\   
In this work, we demonstrate that Dixon-Souriau equations for a charged spinning particle can be recovered from the dimensional reduction of a Papapetrou approach for a 5-dimensional spinning particle, moving in a Kaluza-Klein background.\\
Furthermore, our analysis outlines that, on a 4-dimensional level, an electric dipole contribution arises as linked to the extra-components of the spin tensor. When referred to an elementary particle, the Dixon-Souriau model cannot contain such electric dipole since it would be in contradiction with experiments on fundamental particle physics \cite{Rom01,Re02}. In fact, it would naturally lead to parity and time-reversal violations on a quantum level. We could impose that the electric dipole vanishes by restricting the initial condition on our 5-dimensional Papapetrou approach. However we show that, when the dimensional reduction is performed, the additional dimension can imply an effective parity and time-reversal violation term for spinors. More precisely, we clarify how an electric dipole moment comes out from the Dirac equation on a Riemannian 5-dimensional space-time. We show that it has an order of magnitude fixed by the extra-dimensional length, i.e. negligible in current measurements of particle physics. On the other hand, it is easy to recognized that a similar term does not emerge from the 5-dimensional theory for a scalar or a vector field. Thus, we can conclude that the appropriate initial conditions, for our 5-dimensional study of the Papapetrou equation, must ensure that the electric dipole vanishes identically or has an undetectable magnitude.\\
The structure of the manuscript is as follows: in section 2 we review the Papapetrou approach and the Dixon-Souriau method for the study of a spinning particle and of a charged spinning particle (in an external electro-magnetic field), respectively. Then in section 3, the basic features of the 5-dimensional Kaluza-Klein model, which are at the ground of the later splitting procedure, are illustrated. The geodesics motion of a point-particle on a 5-dimensional Kaluza-Klein space-time is investigated in section 4. The first step to achieve the dimensional reduction of Papapetrou equations is performed in section 5, by splitting the equation for the spin tensor precession. The analysis is concluded in section 6, where we fix equations for the generalized momentum. In section 7 we show how a non-vanishing electric dipole moment can be understood into a Kaluza-Klein framework, even from a quantum point of view, discussing its role for fundamental fields living in a Kaluza-Klein  scenario.  Finally, section 8 is devoted to brief concluding remarks. 

\section{Spinning particle and charged spinning particle in general relativity}
In General Relativity (GR), the matter and the space-time metric are deeply connected by virtue of Einstein equations. Thus, describing the motion of a non-test particle is a rather difficult task, because it demands the solution of the following two coupled systems of differential equations
\begin{eqnarray}
G_{\mu\nu}=\frac{8\pi G}{c^{4}}T^p_{\mu\nu}\qquad\qquad\nabla_{\mu}{T^p}^{\mu\nu}=0\label{1}
\end{eqnarray}
where the former gives Einstein equations and the latter the energy-momentum conservation for a particle with energy-momentum tensor $T^p_{\mu\nu}$. If we introduce an external electro-magnetic field, we have an additional source of curvature and the energy exchange with the matter has to be taken into account. Therefore, due to the non-linearity of the dynamics, we are unable to determine a general solution for the exact problem.\\
A first approximation consists in neglecting the back-reaction of the particle itself on the gravitational field. In this sense, one can simply study the motion on a fixed background and, hence, deal with the second of equations (\ref{1}) only.\\
The standard approach to this test-particle scheme is due to Mathisson \cite{2} and Papapetrou \cite{1}. In particular, Papapetrou treats a vacuum space-time, a part from the narrow tube on which the energy-momentum tensor of the particle does not vanish. Inside this tube, he fixes $x^{\mu}=x^{\mu}(\tau)$ as the coordinates of a curve representing the trajectory of the particle, viewed in the co-moving time $\tau$, while $x^{\mu}$ are taken as coordinates of a generic point within the tube.\\
The calculation procedure is based on evaluating, starting from the conservation of the energy-momentum tensor, the fundamental moments of the particle, through an expansion around the trajectory of the particle (of arbitrary order).\\
At the first order in the multipole expansion, for which all moments except the first one vanish, the Papapetrou method reproduces the dynamics of the structureless test particle, i.e. the geodesic motion.\\
However, the most important result is obtained in the so-called pole-dipole approximation, where the first two moments do not vanish and a characterization of the internal structure of the body is given by the spin tensor
\begin{equation}
S^{\mu\nu}=\int_{\tau}\delta x^{\mu}T^{\nu 0}-\int_{\tau}\delta x^{\nu}T^{\mu 0}
\end{equation} 
being $\delta x^\mu$ the vector connecting the world-line of the particle with the point $x^\mu$ in the tube.\\
In this approximation, Papapetrou provides the following equations for the spinning particle dynamics
\begin{eqnarray}
\left\{\begin{array}{c}\frac{D}{Ds}P^{\mu}=\frac{1}{2}R_{\nu\rho\sigma}^{\phantom1\phantom2\phantom3 \mu} S^{\nu\rho}u^{\sigma}\quad\\
\frac{D}{Ds}S^{\mu\nu}=P^{\mu}u^{\nu}-P^{\nu}u^{\mu}
\label{pe}\quad\end{array}\right..
\end{eqnarray}
with 
\begin{equation}
P^{\mu}=mu^{\mu}-\frac{DS^{\mu\nu}}{Ds}u_{\nu}
\end{equation}
and where $u^\mu$ denotes the 4-velocity of the particle.\\
The first of the two equations above gives the dynamics for the generalized momentum $P^\mu$, while the second one describes the precession of the spin tensor $S^{\mu\nu}$. Thus, we see that rotating particles have non-trivial interactions, through their spin tensor, with the space-time curvature. Even a test particle is expected to deviate from a geodesics curve.\\ 
However, we have to emphasize that the above Papapetrou system (\ref{pe}) is over-determinated, because it contains thirteen unknowns (four components of $P^\mu$, three components of $u^\mu$ and six of $S^{\mu\nu}$), but it consists of ten equations only. Therefore, for its closure, additional conditions are needed. In particular, motivated by the fact that the rotation is spatial, Papapetrou imposes the following ones \cite{3}
\begin{equation}
S^{\mu 0}=0,
\end{equation}
though other two well-grounded choices are available in literature, i.e. the Pirani $S^{\mu\nu}u_{\nu}=0$ and the Tulczyjew $S^{\mu\nu}P_{\nu}=0$ conditions, respectively (we stress that Pirani conditions reduce to Papapetrou ones in the co-moving frame).\\
In this scenario, the introduction of an external electro-magnetic field $F_{\mu\nu}$, coupled to the charged spinning particle, is due to Dixon \cite{5} and Souriau \cite{4}, by a generalization of the Papapetrou dynamics. Here the multipole approximation requires again that the external field does not vary significantly over the extension of the spinning body.\\
More precisely, the analysis of Dixon-Souriau results in the following set of coupled equations 
\begin{eqnarray}
\left\{\begin{array}{c}\frac{D}{Ds}P^{\mu}=\frac{1}{2}R_{\nu\rho\sigma}^{\phantom1\phantom2\phantom3 \mu} S^{\nu\rho}u^{\sigma}+qF^{\mu}_{\phantom1\rho}u^{\rho}+\frac{1}{2}M^{\sigma\nu}\nabla^{\mu}F_{\sigma\nu}\\
\frac{D}{Ds}S^{\mu\nu}=P^{\mu}u^{\nu}-P^{\nu}u^{\mu}-M^{\mu\rho}F_{\rho}^{\phantom1\nu}-M^{\nu\rho}F_{\rho}^{\phantom1\mu}\\
\frac{dq}{ds}=0;\qquad\qquad\qquad\qquad\qquad\qquad\end{array}\right.\label{s3}
\end{eqnarray}
where $q$ denotes the electric charge of the particle and $M^{\mu\nu}$ its electro-magnetic moment; here $P^{\mu}$ and $S^{\mu\nu}$ retain their original meaning, with generalized features, due to the presence of the electro-magnetic field.\\
Furthermore, Souriau outlined how in the case of an elementary particle the electro-magnetic moment has to be taken proportional to the spin tensor (via a scalar function of the electro-magnetic coupling $F^{\mu\nu}S_{\mu\nu}$). In fact, such condition ensures that the electric moment vanishes, in agreement with experimental issues \cite{Rom01,Re02}.

\section{Kaluza-Klein hypothesis}
The geometrization of the electro-magnetic interaction was the issue of the original Kaluza-Klein model \cite{K1,K2}.\\
In this framework, the main point is to recast the gauge boson (the photon) into the space-time metric of a 5-dimensional manifold. In particular, the ansatz on the form of such a metric is the following one 
\begin{equation}
j_{AB}=\left(\begin{array}{cc}g_{\mu\nu}(x^{\rho})+e^{2}k^{2}A_{\mu}(x^{\rho})A_{\nu}(x^{\rho})
& ekA_{\mu}(x^{\rho}) \\
ekA_{\nu}(x^{\rho})
& 1\end{array}\right)\label{5metr}
\end{equation}  
where Greek letters refer to the standard 4-dimensional space-time coordinates ($\mu=0,\ldots,3$) and $A_{\mu}$ to the electro-magnetic potential, while $e$ is the electric charge and $k$ a constant.\\
In view of reproducing General Relativity plus Electrodynamics, the general covariance of the theory is broken down to the 4-dimensional one with the invariance under translations on the fifth dimension. In fact, under this kind of space-time transformations, the behavior of an Abelian gauge boson is reproduced, i.e. its 4-tensor nature and the gauge transformation.\\
These invariance properties are consistent with choosing the 5-dimensional space-time as the direct sum of a 4-dimensional manifold plus a ring of radius $r$. This hypothesis also ensures that the extra-dimension is not detectable at energy scales much less than $1/r$ (in geometric units). In fact, we can Fourier-expand any observable 
\begin{equation}
\label{four}f(x^{\mu};x^{5})=\sum_{n=-\infty}^{+\infty}e^{in\frac{x^{5}}{2\pi r}}f^{n}(x^{\mu})
\end{equation}
and the n-mode $f^{n}$ has the fifth-component of the momentum  of order $\frac{n}{r}$, thus we expect that at low energy with respect to $\frac{1}{r}$ the dynamics in the fifth direction is frozen out to $n=0$ term. So, taking physical quantities which are independent of the variable $x^{5}$ (cylindricity condition), acquires a precise physical meaning for $r\ll 10^{-17}$~cm, and will be at the ground of the below analysis.\\
At this point, starting from the Einstein-Hilbert action in five dimensions
\begin{equation}
\label{k1}S=-\frac{c^{4}}{16\pi {}^{(5)}\!G}\int_{V^{4}\otimes S^{1}}\sqrt{-j}{}^{(5)}\!R d^{4}xdx^{5}
\end{equation} 
and by performing the integration on the additional space, the Kaluza-Klein ``miracle'' happens, i.e. the Maxwell Lagrangian comes out
\begin{equation}
S=-\frac{c^{3}}{16\pi}\frac{2\pi R}{{}^{(5)}\!G}\int_{V^{4}}
\sqrt{-g}\bigg[R+\frac{e^{2}k^{2}}{4}F_{\mu\nu}F_{\rho\sigma}g^{\mu\rho}g^{\nu\sigma}\bigg]d^{4}x.
\end{equation}  
In order to obtain the right Einstein-Maxwell action in four dimensions, the following relations must hold
\begin{equation}
{}^{(5)}\!G=2\pi rG\qquad\qquad k=2\frac{\sqrt{G}}{ec^{2}}
\end{equation}
$G$ being the Newton constant.  

\section{Geodesics in a Kaluza-Klein background}
In the previous section, we outlined how the electro-magnetic potential can be recast into the metric tensor of a proper 5-dimensional space-time manifold. However, a full geometrization of the fundamental interactions cannot be reached till couplings with matter fields are recovered too.\\
A first step in this direction has to deal with the motion of a test particle. Since in 5-dimensions the interaction is geometric, the trajectory of a structureless particle has to be a geodesics, i.e.
\begin{equation}
{}^{(5)}\!u^{A}{}^{(5)}\!\nabla_{A}{}^{(5)}\!u_{B}={}^{(5)}\!u^{A}(\partial_{A}{}^{(5)}\!u_{B}-{}^{(5)}\!\Gamma^{C}_{AB}{}^{(5)}\!u_{C})=0
\label{geo}\end{equation}  
where ${}^{(5)}\!u^{A}=\frac{dx^{A}}{{}^{(5)}\!ds}$ is the 5-velocity, $s$ the affine parameter along the curve, and with capital letters we indicated all space-time coordinates.\\
From the metric (\ref{5metr}), we have the following expression for the 5-dimensional velocity in terms of the four-dimensional one
\begin{eqnarray}
{}^{(5)}\!u_{5}=\frac{d x_{5}}{{}^{(5)}\!d s}=\frac{d x_{5}}{d s}\frac{d s}{{}^{(5)}\!d s}=\frac{1}{\sqrt{1-u_{5}^2}}u_{5}=\alpha u_{5}\label{5u5}\\
{}^{(5)}\!u^{\mu}=\frac{d x^{\mu}}{{}^{(5)}\!d s}=\frac{d x^{\mu}}{d s}\frac{d s}{{}^{(5)}\!d s}=\frac{1}{\sqrt{1-u_{5}^2}}u^{\mu}=\alpha u^{\mu}.\label{5umu}
\end{eqnarray}
We stress that, in order for the above equations to hold, we must have 
\begin{equation}
u_{5}^2<1.\label{u5}
\end{equation}
Further, we can split the geodesic equation (\ref{geo}) with the metric (\ref{5metr}), getting the relations 
\begin{equation}
\frac{d}{ds}u_{5}=0\qquad u^{\nu}\nabla_{\nu}u_{\mu}=\frac{\sqrt{4G}}{c^2}
u^5u^\nu F_{\mu\nu} \label{74}
\end{equation}
by other words, we obtain the equations of motion for a charged particle in the 4-dimensional space-time, as far as the following identification is taken 
\begin{equation}
u_5=\frac{q}{2m\sqrt{G}}.\label{u52}
\end{equation}
Therefore, the dynamics of a charged test particle can be recovered from the free falling in the 5-dimensional scenario. Here the electric charge arises as the fifth component of the momentum.\\
But, the last result stands only for macroscopic objects. In fact from equations (\ref{u5}) and (\ref{u52}) we get the condition
\begin{equation}
\frac{q}{m}<2\sqrt{G}\sim 10^{-4}\frac{C}{kg}.\label{lim}
\end{equation} 
This is an open question of the Kaluza-Klein approach, since elementary particles do not satisfy the above relation (\ref{lim}), and no such condition exists in General Relativity plus Electrodynamics. Neverthless, when referred to a macroscopic body, it stands as the stability condition for a spherical symmetric distribution of charge.\\
The relation (\ref{u52}) also allows to obtain an estimate of the radius $r$. Since the extra-dimension is a circle, the fifth-momentum, i.e. the charge, is related to $r$ as follows
\begin{equation}
p_{5}=\frac{\hbar}{2\pi r}n\qquad n\in N\qquad\Rightarrow\qquad q=\sqrt{G}\frac{\hbar}{\pi r}n.
\end{equation}
According to this approach, we get the quantization of the electric charge as direct consequence of the momentum quantization along a circle. Hence, by imposing the minimum value to be the electron one, we get the following estimate for the radius
\begin{equation}
r=4\pi\sqrt{G}\frac{\hbar}{ec}
\approx2.37\,\,10^{-31}cm.
\end{equation}
The extra-dimension is compactified to scales a few order of magnitude greater than the Planckian one, so that the validity of the cilindricity condition is confirmed at energies available in current experiments.

\section{Splitting of equations for the spin tensor}
The results obtained in the previous section show how the motion of a charged non-rotating object can be inferred from a geodesic one in five dimensions, once the Kaluza-Klein hypothesis are made. As a next step, the dynamics of a spinning particle will be taken into account.\\
To this scope, we consider the 5-dimensional extension of Papapetrou equations constrained by the Pirani condition, i.e. 
\begin{eqnarray}
\left\{\begin{array}{c}\frac{D}{{}^{(5)}\!Ds}{}^{(5)}\!P^{A}=\frac{1}{2}{}^{(5)}\!R_{BCD}^{\phantom1\phantom2\phantom3 A} \Sigma^{BC}{}^{(5)}\!u^{D}\quad\\
\frac{D}{{}^{(5)}\!Ds}\Sigma^{AB}={}^{(5)}\!P^{A}{}^{(5)}\!u^{B}-{}^{(5)}\!P^{B}{}^{(5)}\!u^{A}\\
{}^{(5)}\!P^{A}={}^{(5)}\!m{}^{(5)}\!u^{A}-\frac{D\Sigma^{AB}}{{}^{(5)}\!Ds}{}^{(5)}\!u_{B}\quad\\
\Sigma^{AB}{}^{(5)}\!u_{A}=0\label{pe5}\qquad\qquad\qquad\qquad\end{array}\right.
\end{eqnarray}
and we are going to perform their dimensional reduction.\\
To develop this procedure, we need to identify the meaning of the components of the spin tensor $\Sigma^{AB}$. Under coordinates transformations, proper of a Kaluza-Klein model, we find that $\Sigma^{\mu\nu}$ and $\Sigma_{5\mu}$ behave as 4-dimensional quantities, in particular a tensor $S^{\mu\nu}$ and a vector $S_{\mu}$, respectively.\\ 
First of all, the splitting of the Pirani condition $\Sigma^{AB}{}^{(5)}\!u_{A}=0$ in the case $B=\mu$ gives
\begin{equation}
\alpha(S^{\nu\mu}u_\nu+S^\mu u_5)=0\label{pe5sp}
\end{equation}
while for $B=5$ we get 
\begin{equation}
S^\mu u_\mu=0
\end{equation}
which is contained in the above equation (\ref{pe5sp}), as soon as it is multiplied times $u_\mu$.\\
Hence, we perform the reduction for the expression of the generalized momentum ${}^{(5)}\!P^{A}={}^{(5)}\!m{}^{(5)}\!u^{A}-\frac{D\Sigma^{AB}}{{}^{(5)}\!Ds}{}^{(5)}\!u_{B}$. By using the relations (\ref{5u5}), (\ref{5umu}) and the Christoffel connections obtained from the metric tensor (\ref{5metr}), we get
\begin{eqnarray}
\frac{D\Sigma^{\mu\nu}}{{}^{(5)}\!Ds}=\alpha\bigg[\frac{DS^{\mu\nu}}{Ds}-\frac{1}{2}ekF^{\mu}_{\phantom1\rho}(u^{\rho}S^\nu+S^{\rho\nu}u_5)+\frac{1}{2}ekF^{\nu}_{\phantom1\rho}(u^{\rho}S^\mu+S^{\rho\mu}u_5)\bigg]\\
\frac{D\Sigma^{5\mu}}{{}^{(5)}\!Ds}=\alpha\bigg[\frac{DS^{\mu}}{Ds}-ekA_\tau\frac{DS^{\tau\mu}}{Ds}-\frac{1}{2}ekF_{\rho\nu}u^{\rho}S^{\nu\mu}-\frac{1}{2}ekF^{\mu}_{\phantom1\rho}S^{\rho}u_5-\nonumber\\-\frac{1}{2}e^2k^2[F^{\mu}_{\phantom1\rho}A_\tau(u^{\rho}S^\tau+S^{\rho\tau}u_5)+F_{\rho\nu}A^\nu (u^{\rho}S^{\mu}+S^{\rho\mu}u_5)]\bigg]\\
\frac{D\Sigma_{5\mu}}{{}^{(5)}\!Ds}=\alpha\bigg[\frac{DS_{\mu}}{Ds}+\frac{1}{2}ekF^{\tau}_{\phantom1\rho}u^{\rho}S_{\tau\mu}+\frac{1}{2}ekF^{\tau}_{\phantom1\mu}S_\tau u_5)\bigg]\qquad\qquad\qquad;
\end{eqnarray}
so, with the help of the condition (\ref{pe5sp}), the generalized momentum components can be written in terms of 4-quantities as follows
\begin{eqnarray}
{}^{(5)}\!P^\mu=\alpha^2\widetilde{P}^\mu=\alpha^2\bigg[mu^{\mu}-u_{\nu}\frac{DS^{\mu\nu}}{Ds}-ekF_{\rho\tau}u^{\rho}S^{\tau\mu}u_{5}\bigg]\label{gms}\\
{}^{(5)}\!P_5=\alpha^2\widetilde{P}_{5}=\alpha^2\bigg[mu_{5}-u_{\nu}\frac{DS^{\nu}}{Ds}+ekF_{\rho\nu}u^{\rho}S^{\nu\mu}u_{\mu}\bigg].
\end{eqnarray}
We stress that, after the splitting, we obtain a generalized momentum whose 4-dimensional components depend on the electro-magnetic field $F_{\mu\nu}$. This result is expected, because the interaction of a charged body with the electro-magnetic field contributes to the total amount of its energy-momentum.\\
The next step is to perform the dimensional reduction of the spin tensor precession equations, i.e.
\begin{equation}
\frac{D}{{}^{(5)}\!Ds}\Sigma^{AB}={}^{(5)}\!P^{A}{}^{(5)}\!u^{B}-{}^{(5)}\!P^{B}{}^{(5)}\!u^{A};
\end{equation}
from the above expressions of ${}^{(5)}\!P^\mu$, ${}^{(5)}\!P_5$ and of the derivatives of $\Sigma^{\mu\nu}$ and $\Sigma_{\mu5}$, we obtain, in the case $A=\mu$ and $B=\nu$, the fundamental equation
\begin{equation}  
\frac{DS^{\mu\nu}}{Ds}=\alpha^2[\widetilde{P}^{\mu}u^\nu-\widetilde{P}^{\nu}u^\mu]+\frac{1}{2}ekF^\mu_{\phantom1\rho}(u^\rho S^\nu+u_5 S^{\rho\nu})-\frac{1}{2}ekF^\nu_{\phantom1\rho}(u^\rho S^\mu+u_5 S^{\rho\mu})\label{speqs}
\end{equation}
which describes the dynamics of the spin tensor in presence of the electro-magnetic field.\\
It is worth noting that the dynamics of $\Sigma_{5\mu}$ can be inferred from the equation governing the behavior of $S^{\mu\nu}$. This result ensures that the dynamics of the new degree of freedom $S^\mu$ is determined by the one of the spin 4-tensor and, therefore, its physical meaning, which we will fix below, has no evolutionary character.\\

\section{Reduction of the generalized momentum dynamics}
We can now perform the reduction for equations describing the dynamics of the body. To this end, we need to express the components of the Riemann tensor which, projected on the 5-bein, read as follows

${}^{(5)}\!{R_{(\alpha)(\beta)(\gamma)}}^{(\mu)}={R_{(\alpha)(\beta)(\gamma)}}^{(\mu)}-\frac{1}{4}(F^{(\mu)}_{\phantom1(\alpha)}F_{(\beta)(\gamma)}-F^{(\mu)}_{\phantom1(\beta)}F_{(\gamma)(\alpha)}+2F^{(\mu)}_{\phantom1(\gamma)}F_{(\alpha)(\beta)})$\\
${}^{(5)}\!{R_{(5)(\beta)(\gamma)}}^{(\mu)}=\frac{1}{2}ekF_{(\gamma)}^{\phantom1(\mu)}$\\
${}^{(5)}\!{R_{(5)(\beta)(5)}}^{(\mu)}=\frac{1}{4}e^2k^2F^{(\mu)(\nu)}F_{(\beta)(\nu)}$\\
${}^{(5)}\!{R_{(\alpha)(\beta)(5)}}^{(\mu)}=\frac{1}{4}\nabla^{(\mu)}F_{(\alpha)(\beta)}$\\
${}^{(5)}\!{R_{(\alpha)(\beta)(\gamma)}}^{(5)}=-\frac{1}{2}ek\nabla_{(\gamma)}F_{(\alpha)(\beta)}$\\
${}^{(5)}\!{R_{(\alpha)(\beta)(\gamma)}}^{(5)}=\frac{1}{4}e^2k^2F_{(\nu)(\beta)}F_{(\gamma)}^{\phantom1(\nu)}.$

Hence, we can split the equation for ${}^{(5)}\!P_5$, so getting
\begin{equation}
\frac{D}{Ds}(\alpha^2\widetilde{P}_5)=-\frac{1}{4}ek\frac{D}{Ds}(F_{\alpha\beta}S^{\alpha\beta})
\end{equation} 
therefore the expression above fixes a constant of motion, having the form
\begin{equation}
q=\alpha^{2}\widetilde{P}_{5}+\frac{1}{4}ekF_{\mu\nu}S^{\mu\nu}\label{p3}
\end{equation}
A natural expectation is to recognize $q$ as the electric charge of the moving body, and this assumption is confirmed by the analysis performed below on the equation for ${}^{(5)}\!P^\mu$. We emphasize that, while in the case of a test particle the charge is simply $\widetilde{P}_{5}$, in proper units, a non-vanishing spin tensor implies an additional contribution to its charge.\\
Finally, let us consider the momentum equation equation ${}^{(5)}\!P^\mu$. With the help of relations (\ref{gms}), (\ref{speqs}) and by using the following property of the electro-magnetic field
\begin{equation}
\nabla^{\mu}F^{\nu\rho}+\nabla^{\nu}F^{\rho\mu}+\nabla^{\rho}F^{\mu\nu}=0
\end{equation}
we obtain
\begin{equation}
\frac{D}{Ds}\hat{P}^\mu=\frac{1}{2}{R_{\alpha\beta\gamma}}^\mu S^{\alpha\beta}u^\gamma+ek{F^\mu}_\nu u^\nu\bigg(\alpha^2P_5+\frac{1}{4}F_{\alpha\beta}S^{\alpha\beta}\bigg)+\frac{1}{4}ek\nabla^\mu F^{\nu\gamma}(S_{\nu\gamma}u_5+2S_\gamma u_\nu)\label{momdyn}
\end{equation}
being
\begin{equation}
\hat{P}^\mu=\alpha^{2}P^{\mu}+u_{5}\frac{DS^{\mu}}{Ds}-ekF_{\rho\nu}u^{\rho}S^{\nu\mu}u_{5}+\frac{1}{2}ekF^{\mu}_{\phantom1\rho}S^{\rho}.\label{hat}
\end{equation}
By introducing $\hat{P}^\mu$ also in the equation for the spin tensor, we can restate the relations (\ref{pe5sp}), (\ref{speqs}) and (\ref{momdyn}) so that the full dynamics is described by the following system 
\begin{eqnarray}
\left\{\begin{array}{c}\frac{D}{Ds}\hat{P}^{\mu}=\frac{1}{2}R_{\alpha\beta\gamma}^{\phantom1\phantom2\phantom3\mu}S^{\alpha\beta}u^{\gamma}+qF^{\mu}_{\phantom1\nu}u^{\nu}+\frac{1}{2}\nabla^{\mu}F^{\nu\rho}M_{\nu\rho}\\
\frac{DS^{\mu\nu}}{Ds}=\hat{P}^{\mu}u^{\nu}-\hat{P}^{\nu}u^{\mu}+F^{\mu}_{\phantom1\rho}M^{\rho\nu}-F^{\nu}_{\phantom1\rho}M^{\rho\mu}\\
\hat{P}^{\mu}=\alpha^{2}P^{\mu}+u_{5}\frac{DS^{\mu}}{Ds}-ekF_{\rho\nu}u^{\rho}S^{\nu\mu}u_{5}+\frac{1}{2}ekF^{\mu}_{\phantom1\rho}S^{\rho}\\
S^{\nu\mu}u_{\nu}+S^{\mu}u_{5}=0\label{pe5s}\end{array}\right.,
\end{eqnarray}
where the quantity $M^{\mu\nu}$ admits the expression
\begin{eqnarray}
M^{\mu\nu}=\frac{1}{2}ek(S^{\mu\nu}u_{5}+u^{\mu}S^{\nu}-u^{\nu}S^{\mu})\label{emmom}.
\end{eqnarray}
In the case $S^{\mu}=0$ and $\alpha\sim1$, our results coincide exactly with the dynamics fixed by Souriau for charged spinning particles, being  $M^{\mu\nu}=\lambda S^{\mu\nu}$ the magnetic moment tensor and $\lambda=\frac{1}{2}eku_{5}$ a function of the quantity $F_{\mu\nu}S^{\mu\nu}$ only.\\
On this level, our results hold either in the case of a macroscopic body, either in the case of an elementary particle. However, as far as  $S^{\mu}$ is no longer zero, we are lead to interpret $M^{\mu\nu}$ as the full electro-magnetic moment tensor, including a pure electric dipole component of the form $\frac{1}{2}ek(u^\mu S^{\nu}-u^{\nu}S^{\mu})$. The presence of this term opens the subtle question about the applicability of the scheme here developed to the dynamics of elementary particles endowed with a spin, since, up to now, no experimental evidence for their electric dipole arose.  

\section{Electric dipole moment in a quantum framework}
In the previous section we inferred that, in a 5-dimensional Kaluza-Klein background, the spin tensor extra-components of a moving body describe a non-vanishing electric dipole moment. Within a classical framework, no fundamental symmetry violation arises from this quantity, since it behaves like a polar vector (odd under parity and even under time-reversal), like the electric field, to which it couples.\\
However, in a quantum model, one can show that the expectation value of the electric dipole moment, in an at rest particle state, is proportional to the particle spin (which is even under parity and odd under time-reversal) \cite{Be91}. Therefore, a non vanishing electric dipole moment implies that both parity and time-reversal invariance are broken.\\ 
It is already known a scenario where the introduction of an extra-dimension under Kaluza-Klein assumptions leads to an electric dipole moment, i.e. the spinor field. In fact, we can rewrite the Dirac Lagrangian in a Kaluza-Klein background as follows 
\begin{equation}{\it L}=\frac{i\hbar}{2}[\bar{\psi}\gamma^{(\mu)}(e^{\mu}_{(\mu)}\partial_{\mu}+e^{\mu}_{(\mu)}ekA_{\mu}\partial_{5}-\Gamma_{(\mu)})\psi+\bar{\psi}\gamma^{(5)}(\partial_{5}-\Gamma_{(5)}\psi)]+c.c.
\end{equation} 
being $\gamma_{(5)}=\gamma_5$ the fifth Dirac matrix. Hence, for Riemannian spinorial connections, we obtain the following result 
\begin{equation}
{\it L}=\frac{i\hbar}{2}[\bar{\psi}\gamma^{(\mu)}(e^{\mu}_{(\mu)}\partial_{\mu}+e^{\mu}_{(\mu)}ekA_{\mu}\partial_{5}-{}^{(4)}\!\Gamma_{(\mu)})\psi+\bar{\psi}\gamma^{(5)}\partial_{5}\psi]+\frac{1}{8}F_{\mu\nu}\bar{\psi}\gamma_5[\gamma^\mu;\gamma^\nu]\psi+c.c.
\end{equation} 
the last term being precisely an electric dipole moment contribution.\\ Thirring demonstrated \cite{T72} that for fermions this sort of violations occurs because discrete symmetries have a different definition in a 5-dimensional background, compared to the usual 4-dimensional world.\\In this respect, the emergence of an electric dipole moment in the 5-dimensional Dirac equation is no longer surprising. Furthermore, it is worth noting that such a dipole term is highly suppressed in a Kaluza-Klein scenario \cite{Ic}.\\ On the other hand, for vector particles no electric dipole moment arises as a consequence of the dimensional reduction. The reason is that, while for fermions the $\gamma_5$ matrix appears as one of the $\gamma$ matrix when we extend the Dirac algebra to the 5-dimensional case, for vectors the analogous one, the totally antisymmetric tensor $\epsilon_{\mu\nu\rho\sigma}$, cannot be introduced without violating the 5-parity invariance. This feature requires to impose on vector particles initial conditions such that the components $\Sigma^{5\mu}$ of the spin tensor vanish.\\
Summarizing, when we refer our analysis to a macroscopic body, the electric dipole moment can take an arbitrary value, fixed ab initio by the morphology of the considered system. However, in the case of elementary particles, its value is expected to be not observable (spinors) or to vanish identically (vector bosons). However, in both these two cases, our analysis shows that its dynamics is recognized from the one of the spin 4-tensor.

\section{Conclusions}
In this work, we outlined how the Papapetrou approach and the Kaluza-Klein framework can be matched together, in order to determine the motion of a charge spinning particle. Our results are consistent with those given by Dixon and Souriau in a 4-dimensional background, therefore the geometrization of the electro-magnetic field does not modify the dynamics of the moving object, up to the dipole order.\\
The enlargement of the number of space-time dimensions induces additional degrees of freedom, not only in the metric, but also in the spin tensor. Here, by comparing the two sets of equations (\ref{s3}) and (\ref{pe5s}), proper identifications between 4- and 5-dimensional quantities were provided. However, in this way, the definitions of the electro-magnetic moment and of the generalized momentum itself, get an additional term, proportional to a vector $S^{\mu}$. A natural issue is to identify $S^{\mu}$ with the electric dipole moment of the moving object. This would explain its absence in Souriau treatment for elementary particles. In fact, $S^{\mu}$ is a relic of the spin tensor $\Sigma^{AB}$ after the dimensional splitting is performed. Thus, its presence is deeply connected with the extra-dimension of the space-time, being observable for elementary particles only as a high energy effect. However, the main result of our work is to show that, in the 5-dimensional scenario, an electric dipole comes out naturally from the spin tensor and it does not need to be introduced as an additional degree of freedom.\\
This interpretation of the vector $S^\mu$ as the electric dipole is enforced by the dynamics in the case $S^{\mu}=0$, where we recover the expected proportionality between the spin tensor and the magnetic moment.\\
We have also stressed that a non-vanishing electric dipole moment is a natural feature for spinors in a Kaluza-Klein approach. In fact, once spinors fields are introduced, the parity and time-reversal violations, that the electric dipole induces in a 4-dimensional picture, can be easily accounted, by means of the dimensional reduction procedure (i.e. the fifth Dirac matrix emerges in the 4-dimensional world as the chirality operator $\gamma_5$).\\
A possible extension of this work relays on checking up to which order of approximation, in the moments of the particles, Kaluza-Klein and the Dixon-Souriau points of view are consistent with each other.

\section{Acknowledgments}  
We thank Alessandra Corsi and Orchidea Maria Lecian for their precious help in upgrading the English of the manuscript.

\end{document}